\documentclass[article,onecolumn,amsmath,amssymb,aps,
]{revtex4}

\usepackage{epsfig,bm,epsf,graphics}
\usepackage{graphicx}
\usepackage{dcolumn}
\usepackage{bm}

\begin{document}
\preprint{APS/123-QED}

\title{First-Order Phase Transition in Perovskites Pr$_{0.67}$Sr$_{0.33}$MnO$_{3}$ - Magneto-Caloric Properties - Effect of Multi-Spin Interaction }

\author{Yethreb Essouda$^{1}$\footnote{yethreb.essouda@gmail.com},  Hung T. Diep$^{2}$\footnote{diep@cyu.fr, corresponding author},  Mohamed Ellouze$^{1}$\footnote{mohamed.ellouze@fss.usf.tn}}
\affiliation{%
$^1$Sfax University, Faculty of Sciences of Sfax, LM2EM, B.P. 1171, 3000, Sfax, Tunisia.\\
$^2$  Laboratoire de Physique Th\'eorique et Mod\'elisation,
CY Cergy Paris Universit\'e, CNRS, UMR 8089\\
2, Avenue Adolphe Chauvin, 95302 Cergy-Pontoise Cedex, France.\\
}



\date{\today}


\begin{abstract}

We show by  extensive Monte Carlo simulations that we need a multi-spin interaction in addition to pairwise interactions in order to reproduce the temperature dependence of the experimental magnetization observed in the perovskite compound  Pr$_{0.67}$Sr$_{0.33}$MnO$_{3}$. The multi-spin interaction is introduced in  the Hamiltonian as follows: each spin interacts simultaneously with its four nearest-neighbors.  It does not have the reversal invariance as in a pairwise interaction where reversing the directions of two spins leaves the interaction energy invariant.   As a consequence, it  competes with the pairwise interactions between magnetic ions. The multi-spin interaction allows the sample magnetization $M$ to increase, to decrease or to have a plateau  with increasing $T$.  In this paper we show that $M$ increases with increasing $T$ before making a vertical fall at the transition temperature $T_C$, in contrast to the usual decrease of $M$ with increasing $T$ in most of magnetic systems. This result is in an excellent agreement with the experimental data observed in  Pr$_{0.67}$Sr$_{0.33}$MnO$_{3}$.  Furthermore, we show by the energy histogram taken  at $T_C$ that the transition is clearly of first order.    
We also calculate the magnetic entropy change $|\Delta S_m|$ and the Relative Cooling Power (RCP) by using the set of curves of $M$ obtained under an applied magnetic field $H$ varying from 0 to 5 Tesla across the transition temperature region. We obtain a good agreement with experiments on  $|\Delta S_m|$ and the values of RCP.  This perovskite compound has a good potential in refrigeration  application due to its high RCP.
\vspace{0.5cm}
\begin{description}
\item[PACS numbers: 5.10.Ln;64.30.+t;75.50.Cc]
\end{description}
\end{abstract}

\maketitle


\section{INTRODUCTION}
There is always a challenge for theorists to elaborate models which yield quantitative agreements with experiments in materials science.  On the one hand, this  is because experimental samples are not always "clean" due to various causes such as the presence of impurities, dislocations, domains etc. in addition to the method of sample preparation.  On the other hand, theoretical models are often too simple to describe experimental samples. Of course, if models contain some main ingredients, they can at best get qualitative agreements with experiments.   In magnetism, theories often start with  a Hamiltonian containing pairwise interactions with a spin model (Ising, XY, Heisenberg, Potts, ... spins). The interaction can be of short range, long range or can be of various kinds causing  frustration [1] or topological spin structures (skyrmions) [2]. The nature of the phase transition depends on these elements: it can be of second or first order or unknown critical properties.  Most of transitions due to short-range pairwise interactions in two  or three  dimensions (2D or 3D) are known by the modern theory of phase transitions such as the renormalization group [3] and highly performant numerical simulations such as the histogram techniques and the Wang-Landau method of simulations [4,5]. So far,  in most cases,  the order parameter such as the magnetization $M$ decreases as the temperature $T$ increases and vanishes at the transition temperature $T_C$, except in some frustrated systems where there is no ordering at any $T$ [1], or there is a co-existrence of order and disorder at non-zero $T$ [6,7]. 

In this paper we are interested in a family of  perovskite compounds.   Perovskite compounds have  recently used in many applications ranging from solar cells with  high-efficiency photovoltaics [8,9], lasers [10], light-emitting diodes [11-13], ... to colossal  magnetoresistance [14,15] and magnetic refrigeration devices [ 16].  In what which concerns their magnetic properties,  there is a book edited by E. Dagotto [15] where experimenttal and and theoretical progress have been presented.  The reader is referred to this book for a complete review up to 2003. More recently, it has been shown experimentally [17-23] and theoretically [24,25] that perovskites of manganite family with various doping atoms possess very high Relative Cooling Powers (RCP) which  can be used as clean sources for refrigeration [16]. The so-called "magnetic refrigeration"  is based on the magneto-caloric effects (MCE) discovered by Warburg in 1881 [26] and widely investigated since then (see the review given in [27]).  

In the particular case of the  perovskite compound Pr$_{x}$Sr$_{1_x}$MnO$_{3}$, we have however seen that experimentally $M$ can have a plateau from $T=0$ up to the transition temperature $T_C$ as seen in Pr$_{0.9}$Sr$_{0.1}$MnO$_{3}$ [24 ], or $M$ decreases with increasing $T$ before making a sharp fall at $T_C$ in the strongly bond-disordered Pr$_{0.55}$Sr$_{0.45}$MnO$_{3}$ [25].  Very recently, it has been experimentally observed that  $M$ increases with increasing $T$ before making a sharp fall at $T_C$  in Pr$_{0.67}$Sr$_{0.33}$MnO$_{3}$ which is the subject studied in this paper.  Note that in the theory of phase transitions and critical phenomena, the temperature dependence of the order parameter and the order of the  phase phase transition depends only on  a few parameters: the nature of the interaction (short range, long range, competing interactions,...), the spin model (Ising, XY, Hesenberg, Potts models) and the space dimension [1,3].  In view of the unusual behavior of $M$ mentioned above, we have  introduced a multi-spin interaction in addition to the standard pairwise ones in order to explain these experimental cases [24,25].  This multi-spin interaction  consists in taking  into account the simultaneous interaction of 5 spins at each lattice site. In  the case Pr$_{0.9}$Sr$_{0.1}$MnO$_{3}$ we succeeded to reproduce the magnetization plateau between $T=0$ and $T_C$ experimentally observed. We have also succeeded to reproduce  the experimental magnetization $M$ as a function of $T$ in the case of Pr$_{0.55}$Sr$_{0.45}$MnO$_{3}$. Note that in this  latter case, the sample  is strongly disordered due to the mixing of Mn$_{3+}$ of spin $S=2$ with Mn$_{4+}$ of spin $S=3/2$, in addition to the random dilute positions of Pr$_{3+}$. In such a strongly disordered system, it is rare that the transition is very sharp as experimentally seen in spin glasses. We have shown that with the multi-spin interaction, we could reproduce this result [25].

In this paper, we use again the multi-spin interaction to study the case of the compound  Pr$_{0.67}$Sr$_{0.33}$MnO$_{3}$.  This compound has been experimentally studied. It shows an unusual behavior of $M$ different from the previous cases: $M$ increases as $T$ increases up to $T_C$. before making a very sharp transition [23].  With a fine tuning of parameters, we have reproduced the experimental $M$ versus $T$ as seen below. We have also used the energy histogram method to show evidence that this transition is of first order in agreement with the experimental sharp transition.  We have  calculated the magnetic entropy change and deduced the RCP, both are in agreement with experiments. Note 
that the present compound has a high RCP which is interesting for magnetic refrigeration applications.     

 In section II, we describe our model Hamiltonian and discuss about the multi-spin interaction. The numerical simulation method is also explained. In section III, we show our results and compare to the experimental data on the temperature dependence of $M$.  In section IV, we calculate the magnetic entropy change  $|\Delta S_m|$ under an  applied magnetic field $\mu_0H$ up to 5 Tesla. We deduce the RCP for $\mu_0H$=1, 2, 3, 4  and 5 Tesla. These results are in good agreement with experimental data.  Concluding remarks are given in section V.

\section{Model and Simulation Method}
\subsection{Experimental results of  Pr$_{0.67}$Sr$_{0.33}$MnO$_{3}$ }

Let us describe the structure of Pr$_{0.67}$Sr$_{0.33}$MnO$_{3}$.   It is crystallized in an orthorhombic structure with Pnma space group comprising lattice parameters on three crystalline axes $a=5.4734$Å, $b= 7.7284 $Å and $c=5.4880$ Å. This structure can be modelized as a body-centered tetragonal lattice with the longest axis is the  $b$ crystal axis. The Mn$^{3+}$ and Mn$^{4+}$ occupy the basal planes $(a,c)$, the centered sites are occupied at 67\% by Pr$^{3+}$ and 33\% by Sr.  Since Sr has no spin, the centered sites are magnetically dilute. 
The outermost spins of   Mn$^{3+}$ occupy the orbital 3d$^4$, so its spin is $S=4\times 1/2=2$ by the Hund's rule. The outermost spins of Mn$^{4+}$ occupy 3d$^3$, so its spin is $S=3\times1/2=3/2$. As for Pr$^{3+}$ its outermost electrrons occupy the orbital 4f2, so  its spin is $S=1$ by the Hund's rule. Note that the Mn$^{3+}$ and Mn$^{4+}$ ions occupy randomly the basal sites while  Pr$^{3+}$ occupy random centered sites . We show in Fig. 1 an example of ion distribution in a  bct cell.  

\begin{figure}[ht]
\centering
\includegraphics[width=8cm,angle=0]{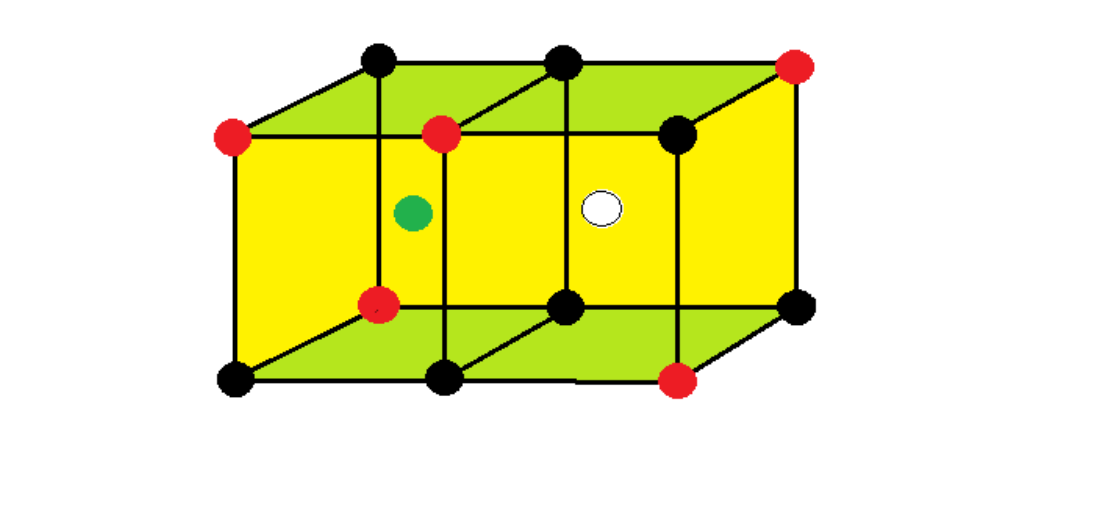}
\caption{An example of ion distribution on two neighboring bct lattice cells: black, red and green circles represent Mn$^{3+}$, Mn$^{4+}$ and Pr$^{3+}$, respectively, white circle is a Sr site with spin zero.  }
\label{fig1}
\end{figure}

Experiments have been performed on this compound [23]. The data show that $M$ in zero-field cooling slighly increases with $T$ and makes a sharp fall at $T_C=291$ K.  As said in the Introduction, this behavior is strikingly unusual.  Note that the compounds with the percentage of Pr close to 0.67 show different temperature dependence of $M$:   Pr$_{0.63}$Sr$_{0.37}$MnO$_{3}$ [19 ],  Pr$_{0.60}$Sr$_{0.40}$MnO$_{3}$ [17 ],  Pr$_{0.7}$Sr$_{0.3}$MnO$_{3}$ [17 ]. Ref. [18] shows curves of $M(T)$ in various applied fields, so there is strictly speaking no phase transition for ferromagnets in an applied field. Refs. [17,22] show a tendency of $M$ increasing with increasing $T$ as in Ref. [23] but the transition is not sharp due to a small applied field.  We think that the  work in [23] is more recent with probably higher performant experimental techniques. We will focus on the results of this work to  elaborate our theoretical model.

\subsection{Model}
The  Pr$_x$Sr$_{1-x}$MnO3 compounds have complicated structures due to the random mixing of spins $S=2$ of  Mn$^{3+}$, $S=3/2$ of  Mn$^{4+}$ and $S=1$ of  Pr$^{3+}$ in addition to the zero-spin Sr sites . These are strongly disordered systems of mixed spins, specially when $x$ is between 0.4 and  0.8.  To our knowledge, there are  so far no theoretical works taking into account the above-mentioned factors, except our two previous [24,25].  We have tried various pairwise interactions and various spin models but none gives sharp first-order-like fall of $M$ at $T_C$.  In statistical physics, few pairwise interactions give rise to a first-order transition: we can mention what is known: frustrated systems in 3D, $q$-state Potts models in 2D with $q> 4$ or in 3D with $q \geq 3$ in 3D. The perowskite compounds we study are not frustrated because, in spite of the mixed spins,  the ferromagnetic interaction between    Mn$^{3+}$ and Mn$^{4+}$ due to double exchange  interaction via intercalated oxygen ions [28-31]  is dominant, leading to a ferromagnetic ordering. So the sharp fall of $M$ at $T_C$ in these compounds may be due to other types of interaction. We have proposed in previous works [24,25] a multi-spin interaction which reproduce not only the sharp fall of $M$ at $T_C$ but also its temperature dependence from 0 to $T_C$. As will be shown below, the multi-spin  interaction creates spin fluctuations under control by tuning its strength with respect to the pairwise interactions.

There are 6 kinds of  pairwise interactions in addition to the multi-spin interaction:
\indent $J_{1}$: Interaction coupling of a Mn$^{3+}$ ion with a NN Mn$^{3+}$ ion,\\ 
\indent $J_{2}$: Interaction coupling of a Mn$^{3+}$ ion with a NN Mn$^{4+}$ ion,\\  
\indent $J_{3}$: Interaction coupling of a Mn$^{4+}$ ion with a NN Mn$^{4+}$ ion,\\  
\indent $J_{4}$: Interaction coupling of a Pr ion with a Mn$^{3+}$ ion,\\
\indent $J_{5}$: Interaction coupling of a Pr ion with a Mn$^{4+}$ ion\\
\indent $J_{6}$: Interaction coupling between two Pr ions on the adjacent bct units.\\

The pairwise Hamiltonian is written as

\begin{equation}\label{pairwise}
 {\cal H}_p = -\sum_{<i,j>}J_{ij}\mathbf S_{i}\cdot \mathbf S_{j}-\mu_0H\sum_{<i>}S_{i}
\end{equation}
where $\mathbf S_i$ is the spin at the lattice site $i$, $\sum_{<i,j>}$ is made over spin pairs coupled through
the exchange interaction $J_{ij}$.  $H$ is a magnetic field applied along the $z$ axis. Depending on the kinds of the ions at lattice sites $i$ and $j$, we have $J_{ij}$ equal to one of the above
six kinds of interaction. Using only these interactions and with  spin models Ising and Heisenberg, we did not succeed to reproduce the curve $M$ versus $T$ without a multi-spin interaction, as shown in [24,25].

At this stage, let us note  that the pairwise interaction used in magnetism comes from model Hamiltonians of statistical physics. All of them, except the Heisenberg case, are introduced by "hand", but their validities are verified by many experiments: we can mention the Ising model [32] and various Potts models [33] among others.  The Heisenberg case is exceptional: starting from the overlap of two neighboring orbitals, and using their spin-dependent wave functions in the Hartree-Fock approximation it was shown  that the second-order perturbation gives rise to the coupling between the spins of neighbor atoms (the reader is referred to the detailed demonstration given on pp. 53-55 of Ref. [34]).   It is however intuitively, the interaction between a spin with all of its neighbors should be simultaneous. Strictly speaking it is not the sum of interacting pairs.  However, mathematically it is not possible to demonstrate a multi-spin  interaction from the first principles. It is nevertheless, starting from vertex model   one can exactly show that  Ising models with $m$-spin interactions ($m> 2$) can be derived from the square-lattice eight-vertex
model [35] which can be mapped onto an Ising model with two- and four-spin
interactions by Wu [36] and by Kadanoff and Wegner [37]. Let us mention also the Ashkin-Teller model [38] which was also reformulated as an Ising model on the square lattice with two- and four-spin
interactions by Fan [39]. Note in passing that the Ising model with three-spin interactions on the triangular
lattice was solved by Baxter and Wu [40,41]. A recent work on the $m$-spin interaction in 1D has been studied by Turban [42]. The reader is referred to this work for  mathematical  details.
So, the formulation   of various multi-spin interactions is not new, but their effects are not widely studied, although there is a renewed interest seen in some very recent works [42-44].  On this aspect, our multi-spin interaction used to explain various unusual behaviors of $M$ in real materials [24,25] is the first attempt to enrich the family of magnetic interactions in real materials. 

\begin{figure}[ht]
\centering
\includegraphics[width=6cm,angle=0]{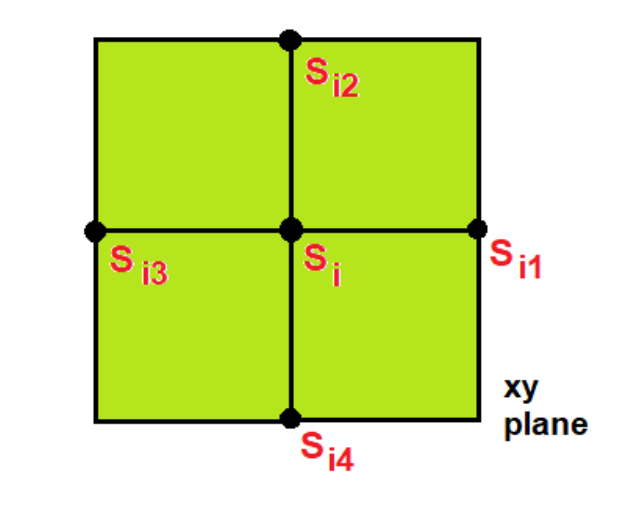}
\caption{Spin of the Mn ion at the site $i$ interacts simultaneously with four Mn spins at nearest sites $i1$, $i2$, $i3$ and $i4$ in the multi-spin interaction given by Eq. (\ref{multi}). See text for comments.  }
\label{fig2}
\end{figure}

The multi-spin interaction is written as

\begin{equation}\label{multi}
{\cal H}_m = - K\sum_{i} S_{i} S_{i1} S_{i2} S_{i3} S_{i4}
\end{equation}
where $K$ is the interaction strength and the sum runs over all Mn sites $i$. The spins $S_i$ interacts simultaneously with its four nearest neighbors (NN) $ S_{i1}, \ S_{i2},\  S_{i3}$ and $ S_{i4}$  on the $xy$ plane (see Fig. \ref{fig2}). 

Note that the Ising pairwise interaction obeys the reversal symmetry, i. e. the properties of the system are invariant by the operation ($S_i\rightarrow -S_i$,$S_j\rightarrow -S_j$). Also, the properties of  the  system do not change under  the Mattis transformation ($S_i\rightarrow -S_i$,$J_{ij}\rightarrow -J_{ij}$). This transformation just changes  the feromagnetic ordering to antiferromagnetic ordering, but the  transition temperature and the critical exponents remain the same (see Ref. [34], p. 143).  In contrast to pairwise interactions, the multi-spin interaction given above does not have the "overall" reversal symmetry (reversing all spins) because the energy changes its sign. However, if an even number (2 or 4) of spins changes their  sign, the energy remains invariant. This 
property is interesting because it creates a large degeneracy which gives rise to first-order transitions as seen in $q$-state Potts models mentioned above. We return to this point below.

Let us note that in our model the spin of amplitude $S$ is a multi-state Ising spin, taking its values among $-S,-S+1,..,S-1,S$ ($2S+1$ values). This is contrast to the standard Ising model $\pm S$. The multi-state Ising spin allows for a smooth variation of energy in the spin updating.

\subsection{Method of simulation}
We  first use the  Metropolis algorithm [45]  to perform simulations. The sample sizes are 30$^3$ bct cells, namely $2\times30^3$ lattice sites. The total number of spins is $N$. We generate a random spin distribution on the lattice of    Pr$_x$Sr$_{1-x}$Mn$_x^{3+}$Mn$_{1-x}^{4+}$ with $x=0.67$  We use periodic boundary conditions in all directions to reduce surface effects.  The first $10^5$ MC steps are used to equilibrate the system, and the averaging of physical quantities is  taken during the following $10^5$ MC steps.  The calculated physical quantities are:  the average internal energy $E$ per spin, the specific heat $C_V$ per spin, the magnetization $M$ and the susceptibility $\chi$ per spin. They are defined by
\begin{eqnarray}
\langle E \rangle &=&\frac{1}{N}\langle ({\cal H}_p+{\cal H}_m)\rangle \\
C_V&=&\frac{1}{k_BT^2}[\langle E^2\rangle - \langle E\rangle ^2]\\
M&=&g\mu_B (\langle M_{1}\rangle +\langle M_{2}\rangle+\langle M_{3}\rangle) \label{mtot}\\
\chi&=&\frac{1}{k_BT} [\langle M^2\rangle -\langle M\rangle^2]
\end{eqnarray}
where $M_{\ell}$ ($\ell=1,2,3$) is the average of spins of kind $\ell$:  $\ell=1$ for Mn$^{3+}$, $\ell=2$ for Mn$^{4+}$, and $\ell=3$ for Pr$^{3+}$, defined as 

\begin{equation}
\langle M_{\ell}\rangle =\frac{1}{N_{\ell}}\langle\sum_{i\in \ell}S_i\rangle \label{msub}
\end{equation}
where $N_{\ell}$ is the number of spins of kind $\ell$.  Note that the  Land\'e factor $g$ in (\ref{mtot}) should be understood as an "effective" $g$ since it concerns three types de spin.

Let us emphasize that in  the updating of spins, we use the over-relaxation method by Creutz [46,47] which is known to accelerate the convergency to equilibrium   [48]. The reader is referred to [25] for more details of the implementation.  In short, at a given spin $S_i$, we calculate its interaction energy using Eqs. (\ref{pairwise}) and (\ref{multi}). We take  a new value of $S_i$ among its ($2S+1$) values, we calculate its new energy. This new state is accepted or not using the Metropolis algorithm.  Now, instead of taking another spin to proceed, we stay with the spin $S_i$ and try to update it several times: this is called over-relaxation updating by Creutz [46,47] which optimizes the equilibrium state. The reproductivity of the results  is checked by using independent runs with different random spin distributions. 

Before showing our results, let us emphasize that for the purpose of the present paper, we do not need to use high-performance MC techniques such as the multi-histogram method [4] (see an example in [49]) or the Wang-Landau algorithm [5] (see an example in [50]).  We shall show below however a simpler energy histogram technique which indicates evidence of the first-order character of the transition observed the the present compound. 

\section{Results - Comparison with Experiment } \label{results}

The temperature dependence of $M$ is the main experimental data which is used for our modeling. The energy is not experimentally accessible. The specific heat and the susceptibily are not experimentally available. 

To fit with the experimental $M$, the best set of parameters we found are

\begin{eqnarray}
J_1&=&-0.1J,\  J_2= +1.95J,\  J_3=-0.1J,\  J_4= -0.30J, \ J_5= -0.25J, \ J_6= +0.05J \label{val1}\\
C&=&0.5  \label{val2}\\
K&=&1.42  \label{val3}
\end{eqnarray} 
where $J$ is the energy unit for the simulations. Its real value is obtained when we fit the simulation $M(T)$ curve with the experimental $M(T)$.   $C$ is the reduction coefficient applied  for interactions along the $z$ axis whose lattice constant is much longer than those in the $xy$ plane.

\begin{figure}[ht]
\centering
\includegraphics[width=6.3cm,angle=0]{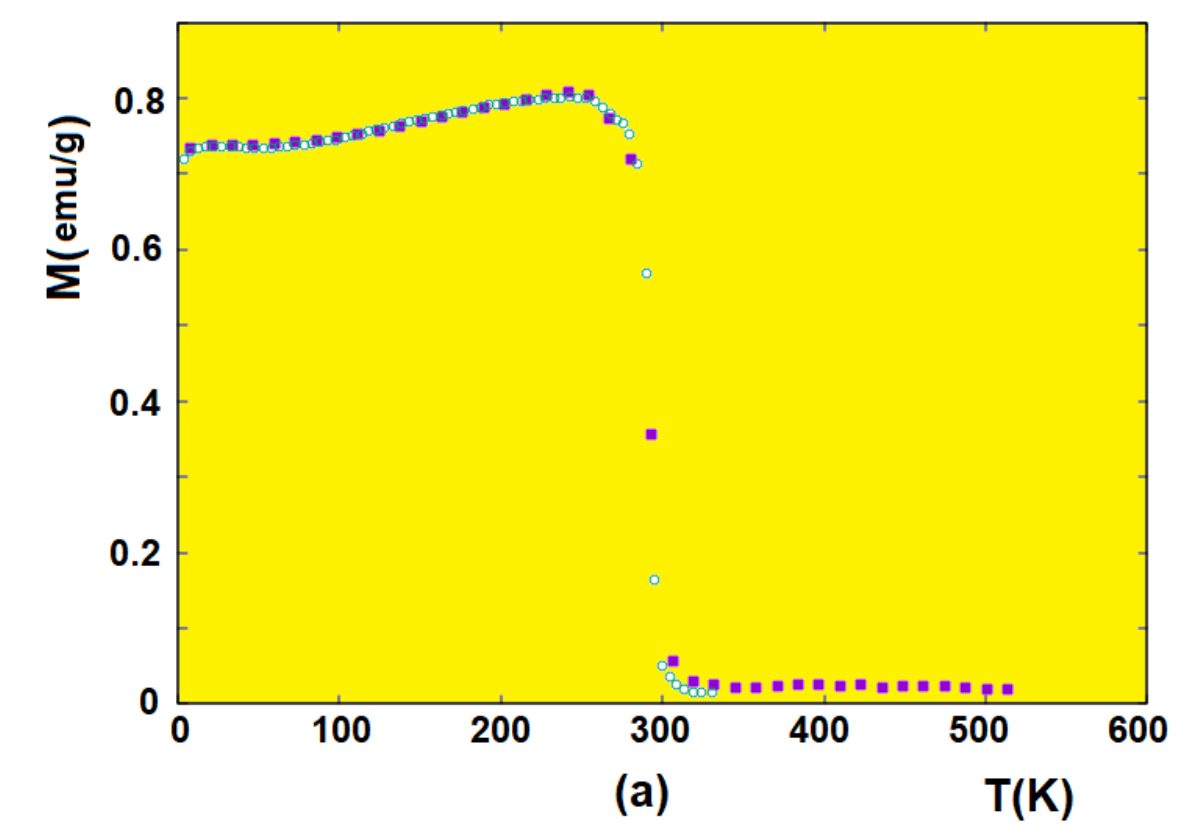}
\includegraphics[width=6cm,angle=0]{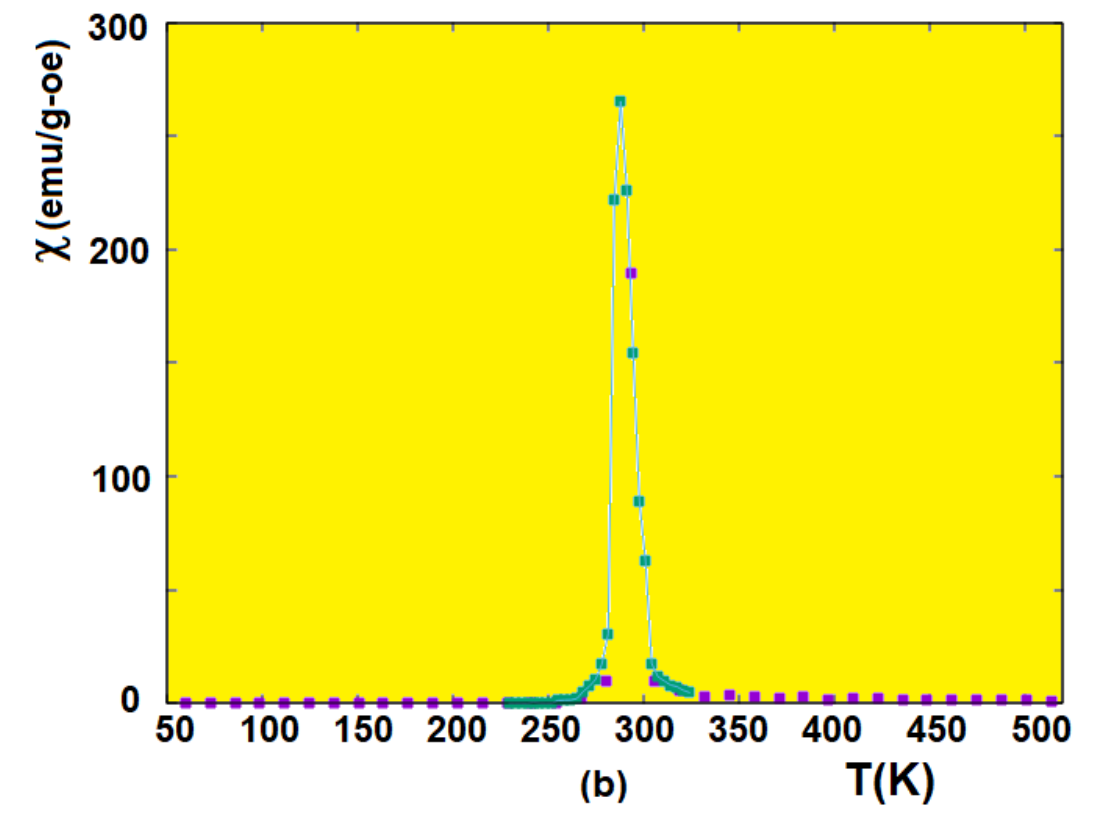}
\caption{(a) MC result (violet squares) and experimental magnetization (green void circles) versus  temperature $T$ in Kelvin, are shown for comparison,
(b) Magnetic susceptibility   versus  temperature $T$. See text for comments.  }
\label{mag}
\end{figure}

%
Using the above interaction values we obtained the MC transition temperature $T_C(MC)\simeq 3.684 J$.  
To find the value of $J$, we use the equality $T_C=291 K=3.684 J$ since the transition temperature is proportional to the energy unit. We have $J\simeq 79$ Kelvin.  This $J$ is however an effective interaction since there are several kinds of interaction. Let us show the MC curve and the experimental magnetization  $M$  in Fig. \ref{mag}a.  By fitting the value of MC magnetization at $T=0$ with experimental value of Ref. [23], one obtains $g\mu_B\simeq 0.74$ emu/g.  Figure \ref{mag}b shows the MC result for the susceptibility $\chi$ vs $T$  (two colors are two independent runs in order to have more data near $T_C$).  

Several remarks are in order:
\begin{itemize}
\item The MC data coincide with the experimental magnetizatio for the  whole temperature range from 0 to $T_C$,
\item The magnetization increase slighly with increasing $T$ up to $T_C$, this unusual behavior was also experimentally observed in early works [17,22] but the transition was not sharp as in the recent work [23],
\item The fall of $M$ at $T_C$ is vertical, suggesting a first-order transition.  We will show evidence of this below. 
\item The peak of $\chi$ is very sharp.
\end{itemize}

Knowing $J$, we can calculate the interactions $J_1,\ J_2, ...,J_6$ in real unit. For example the dominant interaction $J_2=1.95J=154.04$ Kelvin, and $J_1=-0.1J=-7.9$ Kelvin etc.
The MC result gives $E_0(MC)=-14.1$ in unit of $k_BJ$. 
The GS energy in real unit (meV) is
\begin{eqnarray}
E_0( GS)&=&-14.1\times 79 \times k_B\nonumber \\
&=& -1537.90 \times 10^{-23}\ \  {\mbox {Joules}}\nonumber \\
&=& -1537.90 \times 10^{-23}\times 6.242x10^18 \ \ {\mbox {eV}} \nonumber \\
&=&- 9599.60 \times 10^{-5}\ \  {\mbox {eV}}\nonumber\\
&\simeq &-96 \ \  {\mbox {meV}} \label{EGS}
\end{eqnarray}
Now, if we view the system as an "effective" ferromagnet with an effective exchange interaction $J_{eff}$,  we can use the following expression for the ground-state (GS) energy

\begin{equation}
E_0(GS)=-0.5\times Z\times S_{eff}^2 \times  J_{eff}\label{MCE0}
\end{equation}
where $Z=6$Mn+$8\times 0.67$Pr=11.36  is the effective coordination number at a Mn lattice spin, $S_{eff}$ is the effective spin length given by 
$S_{eff}=(0.67\times 2+0.33\times 1.5+0.67\times 1)(0.67+0.33+0.67)\simeq1.503$. 
\begin{eqnarray}
 J_{eff} &\simeq&-96/[-0.5\times11.36\times1.503^2] \ \ {\mbox {meV}}\nonumber\\
 &=& + 7.4817\ \  {\mbox {meV}} \nonumber\\
 &=& 74.8 \ \ {\mbox K}
\end{eqnarray}
Note that this value of the effective exchange interaction, very close to $J$ (=79 K),
 is in the order of magnitude of interactions in magnetic materials with  $T_C$ at the room temperature [51,52].

\begin{figure}[ht]
\centering
\includegraphics[width=8cm,angle=0]{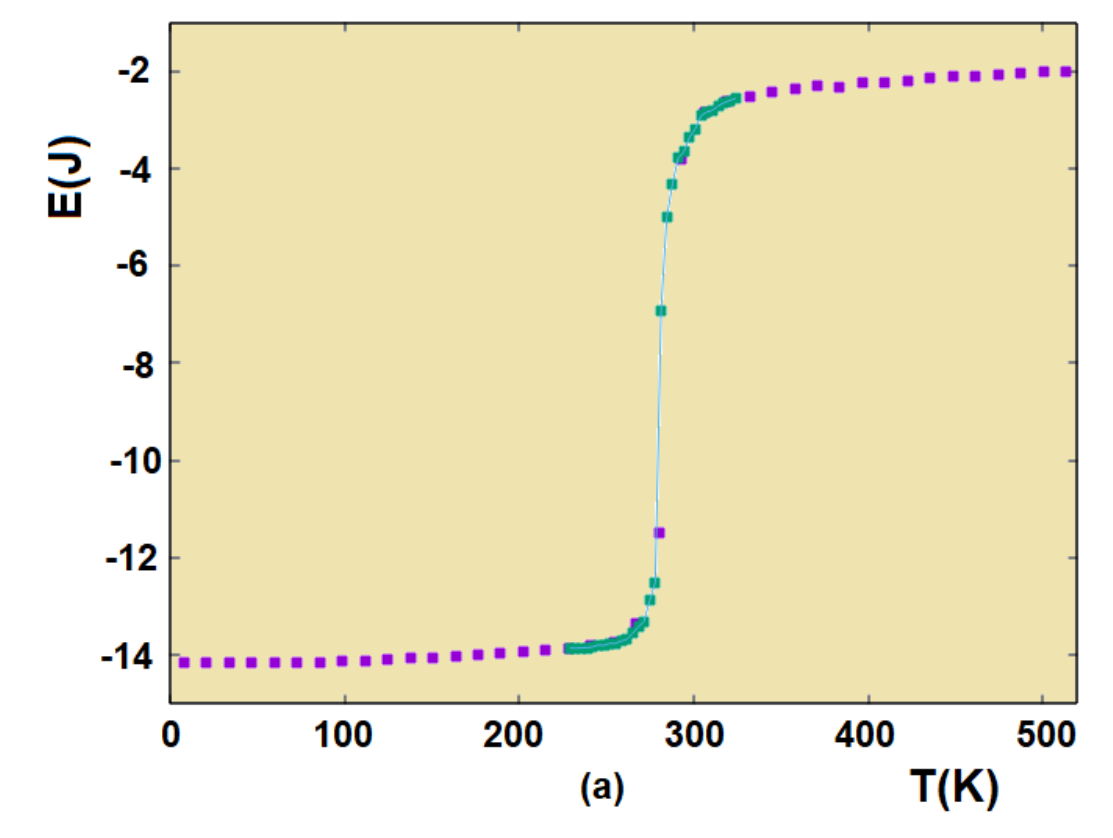}
\includegraphics[width=8cm,angle=0]{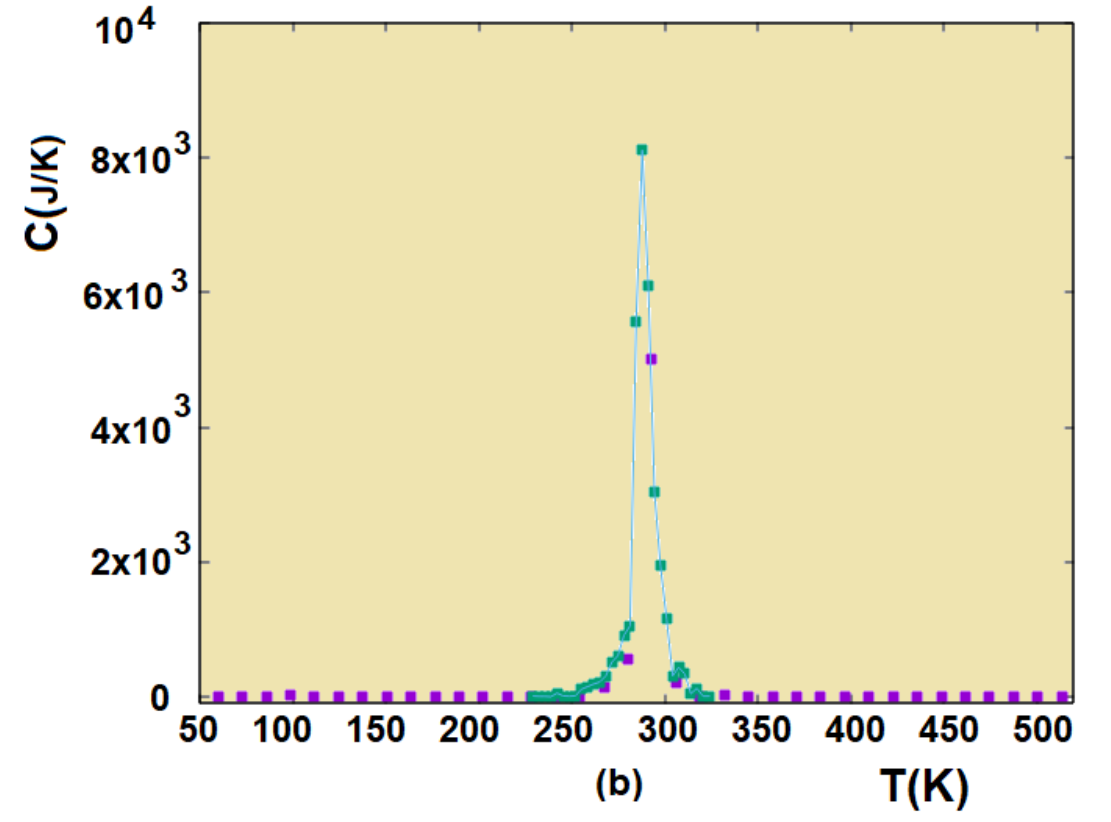}
\caption{(a) Internal energy per spin versus  temperature $T$ in Kelvin, (b) Specific heat   versus $T$. Line is guide to the eye.}
\label{fig4}
\end{figure}

Let us show the internal energy per spin $E$ and the specific heat $C_V$ in Fig. \ref{fig4}.  Two colors correspnd to two runs to have more data around $T_C$. Some remarks are in order:

\begin{itemize}
\item The GS energy is 14.1 in unit of $k_BJ$ in the figure which is -96 meV [see (\ref{EGS})],
\item $E$ has a very stiff slope at $T_C$.  We will show below by energy histograms that in fact $E$ is discontinuous at $T_C$,
\item The peak of $C_V$ is extremely high as a consequence of strong fluctuations due to the muli-spin interaction. We will give a discussion below.
\end{itemize}

\begin{figure}[ht]
\vspace{0.5cm}
\centering
\includegraphics[width=8cm,angle=0]{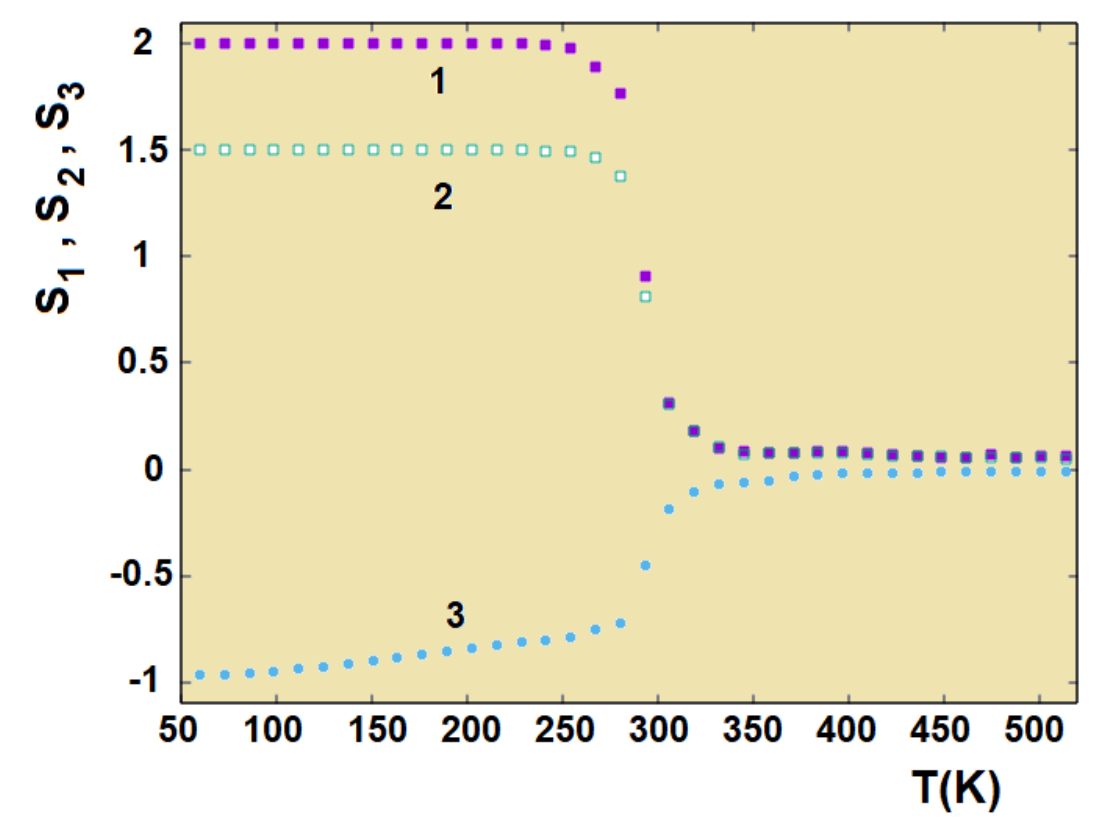}
\caption{ Curves 1, 2 and 3 show the sublattice $z$ components $S_1$, $S_2$ and $S_3$  of  Mn$^{3+}$, Mn$^{4+}$ and Pr$^{3+}$, respectively, versus  temperature $T$.  }
\label{fig5}
\end{figure}

We show now in Fig. \ref{fig5} the $z$ components of three sublattice magnetic ions $\langle S_i \rangle\ (i=1,2,3)$.   We see here that the Mn$^{3+}$ ($S_1$) and  Mn$^{4+}$ ($S_2$) have the same sign, namely they order ferromagnetically, while Pr$^{3+}$ ions ($S_3$)  are ordered antiferromagnetically with the Mn ions.  These data are MC results, they were not available by experiments.  Note that all curves fall sharply at $T_C$.  This results in the vertical fall of $M$ shown in Fig. \ref{mag}a. 

\subsection{Energy Histograms: Evidence of First-Order Transition}
In the theory of phase transitions and critical phenomena, a transition of first order is characterized by a discontinuity of the first derivatives of the free energy (see chapter 7 "Phase Transitions" in Ref. [34]). In our problem, it is $M$ and $E$ which should be discontinuous at $T_C$.  Take $E$ for example. At $T_C$, theoretically there is a co-existence of the ordered phase of energy $E_1$ and the disordered phase of energy $E_2$.  The distance $E_2-E_1$ is called "latent heat" in first-order transitions.
Let us expand a little bit how in MC simulations a first-order character is observed:  due to the  updating dynamics, the co-existence of two phases can take place in two ways: i) at a time $t$, the two phases spatially coexist (half space for each phase for example), ii)  at time $t$ one of the phases occupy the whole sample and this lasts for a lapse of time, it changes to the other phase during a next lapse of time, and repeats this cycle.  For the first scenario, if one averages the energy on the sample at $t$, we find the average value $\bar E=(E_1+E_2)/2$ which is located at the middle  of the vertical slope at $T_C$, one can have a  wrong feeling that the energy is not discontinuous. In fact, it is. For the second scenario, if we again average the energy of the system during a long time, as we do in MC simulations, the time-averaged energy is also      
 $\bar E=(E_1+E_2)/2$ which is located in the vertical slope of $E$ at $T_C$ (we suppose the two lapses of time are equal).  So, how to see the discontinuity of $E$ at $T_C$? The answer is we have to record the energy at $t$ for a very long time and make a histogram $h(E(t))$.  The second scenario will be  seen by a double-peak structure of $h(E(t))$, one peak is centered at $E_1$ and the other peak centered at $E_2$. 

\begin{figure}[ht]
\centering
\includegraphics[width=7cm,angle=0]{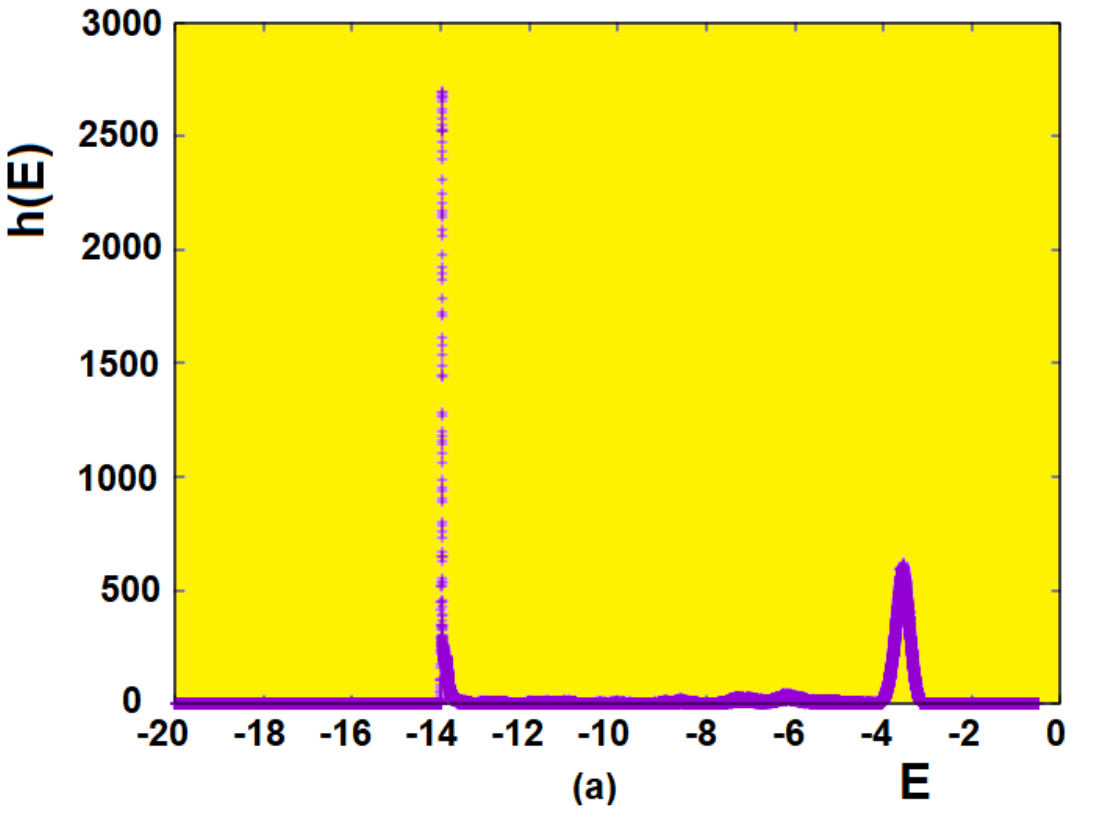}
\includegraphics[width=7cm,angle=0]{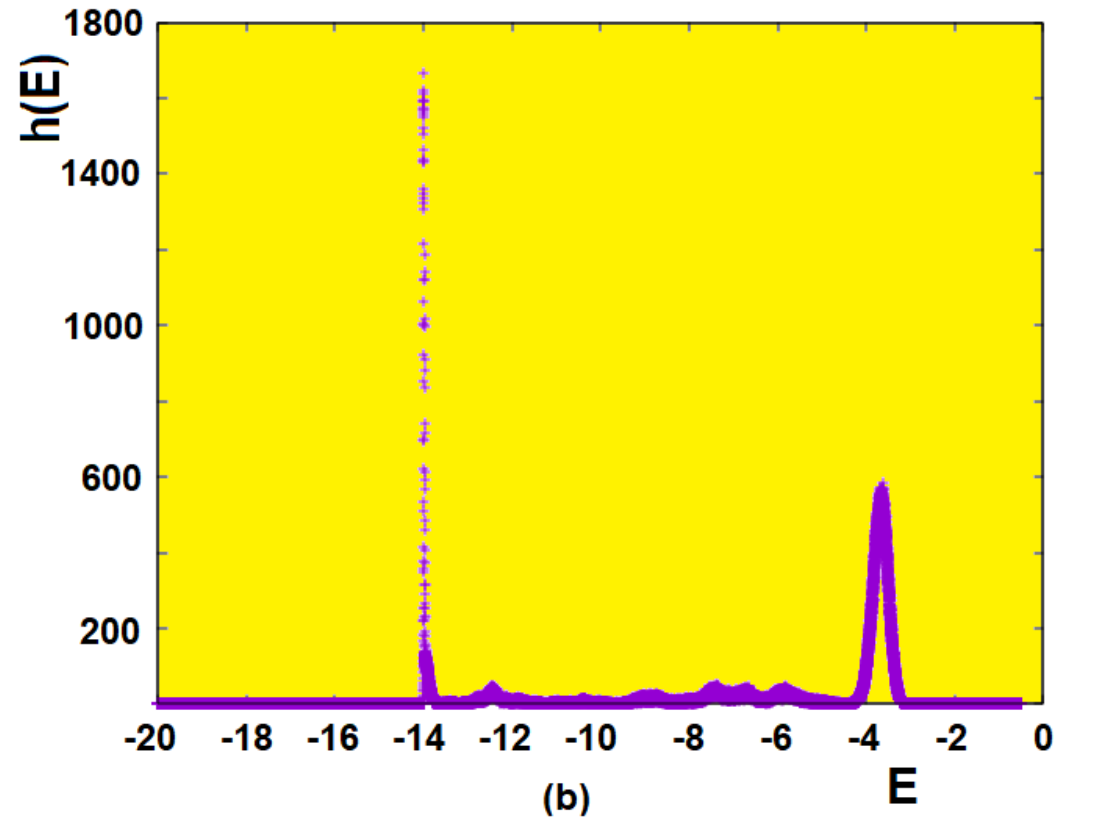}
\includegraphics[width=7cm,angle=0]{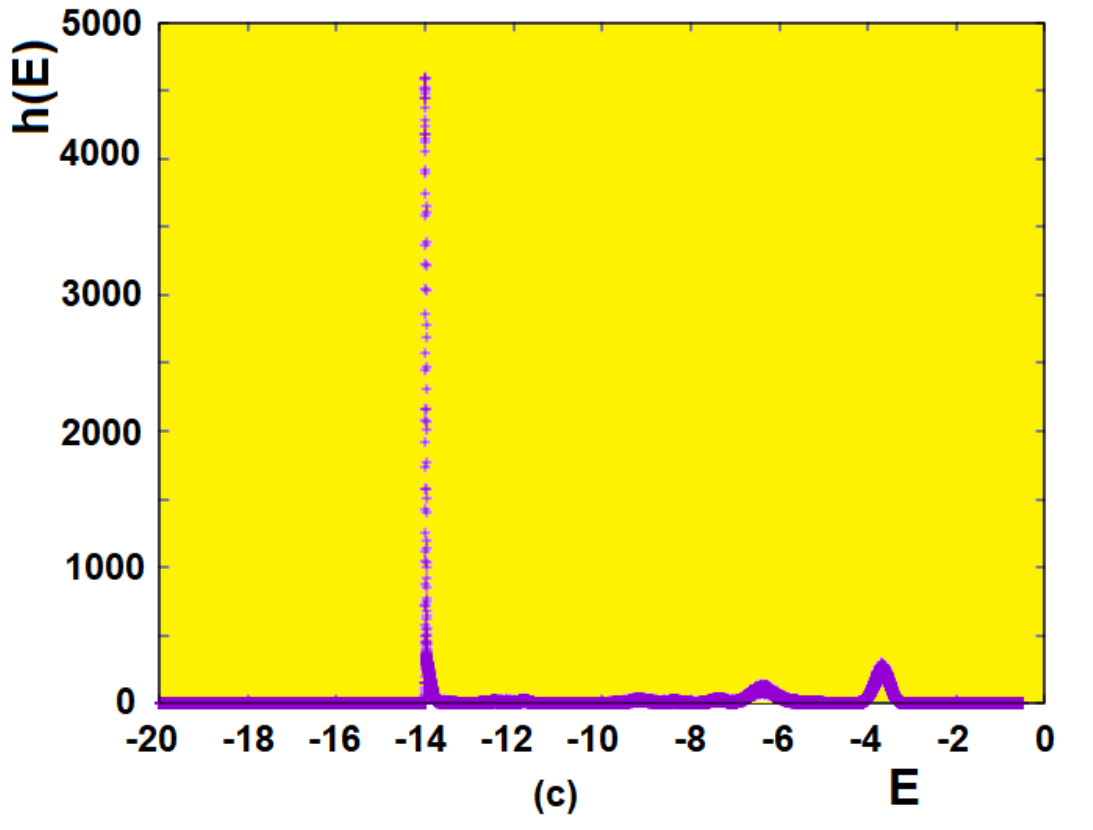}
\includegraphics[width=7cm,angle=0]{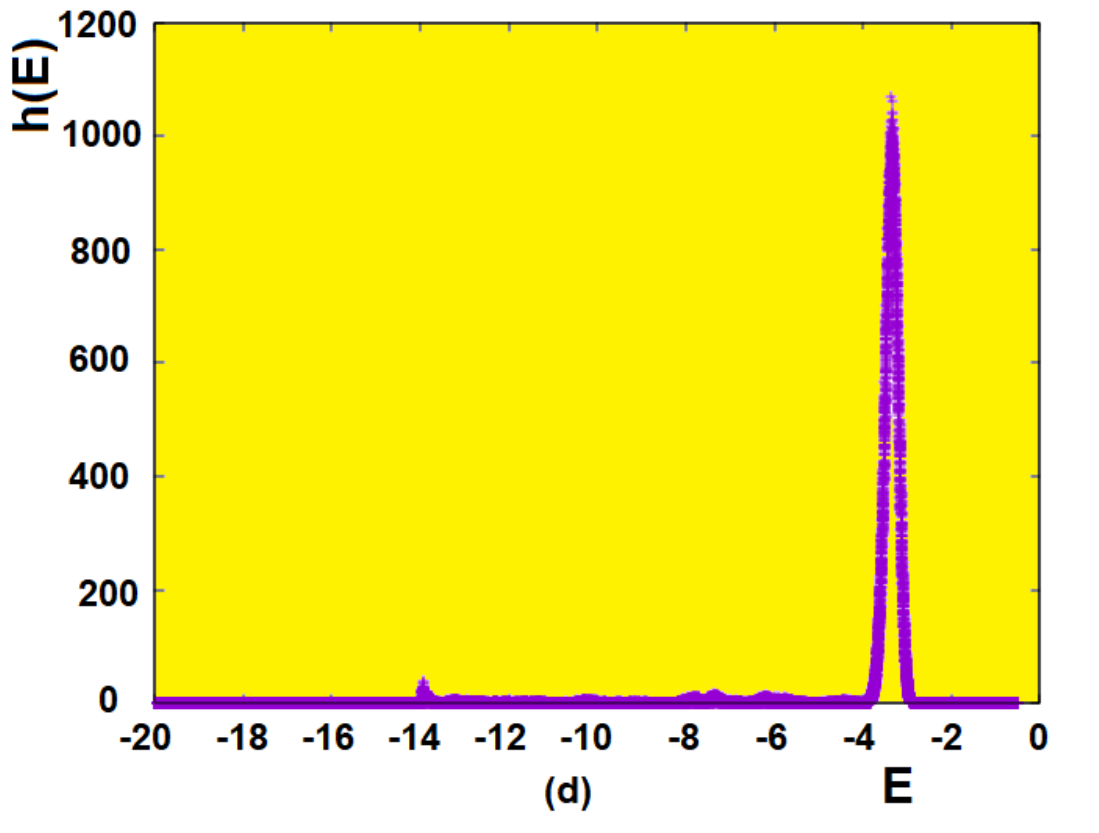}
\caption{ Energy histograms $h(E)$ at $T$=291, 290.90, 282.50, 308.46 K taken over $10^6$ MC steps.  See text for comments.}
\label{histo1}
\end{figure}

We have realized the energy histogram at and near $T_C$. We show in Fig.  \ref{histo1} this histogram for several temperatures at and close to $T_C$: Figs.  \ref{histo1}a and
Fig.  \ref{histo1}b are for $T$=291 K and 209.90 K, respectively, namely at $\simeq T_C$. Figures  \ref{histo1}c and  \ref{histo1}d are taken at $T=$282.50 K below $T_C$ and $T=308.46$ K above $T_C$. 
As discussed at the beginning of this subsection, Fig.  \ref{histo1}a and Fig.  \ref{histo1}b correspond to the second scenario in which the system goes between the ordered (left peak) and disordered (right peak) phases. 
As said above, if we take the energy average of Fig.  \ref{histo1}a, we have $<E>=-6.42686$ which is located in the middle of the energy gap. For  Fig.  \ref{histo1}b, we have a slightly different $<E>=-5.89695$ located also in the energy gap. The points in the vertical slope of Fig.  \ref{fig4}a are results of the average: in fact the system does not stay at the averaged points, but at $E_1$ and $E_2$ as shown by the double peaks in these figures. Below and above $T_C$, we have only one peak, the averaged energies are  near the positions of the peaks, taking into account the tiny peaks.

To conclude this section, we emphasize on two points:

i) the multi-spin interaction allows for an excellent agreement on $M(T)$ between our model and experiments,

ii) the energy histogram technique clearly shows evidence that the transition in the present perovskite compound is of first order.

\section   {Magnetocaloric Effect - Magnetic Entropy Change}

In order to calculate the magnetic entropy change, we apply a magnetic field $H_i\in (0,...,H)$ on the sample at a given $T$. We calculate the magnetization $M(T,H)$. This is shown in Fig. \ref{fig7}. These results are to be compared with the experimental data taken from [23] of sample P1200 shown in Fig. {\ref{fig8}.  We see that the agreement between experiment and simulation is good except at very low $H$ where  simulations give higher magnetizations at temperatures around $T_C$ : experimental curves have stronger slopes, while numerical curves  attain horizontal slopes at lower fields in this  temperature range. This is expected since simulation samples are exempt of defects, domains,... which are more or less present in experimental samples. These defects prevent an increase of $M$ at low $H$ in experiments, but they do not resist to $H$ at stronger values.

\begin{figure}[ht]
\centering
\includegraphics[width=10cm,angle=0]{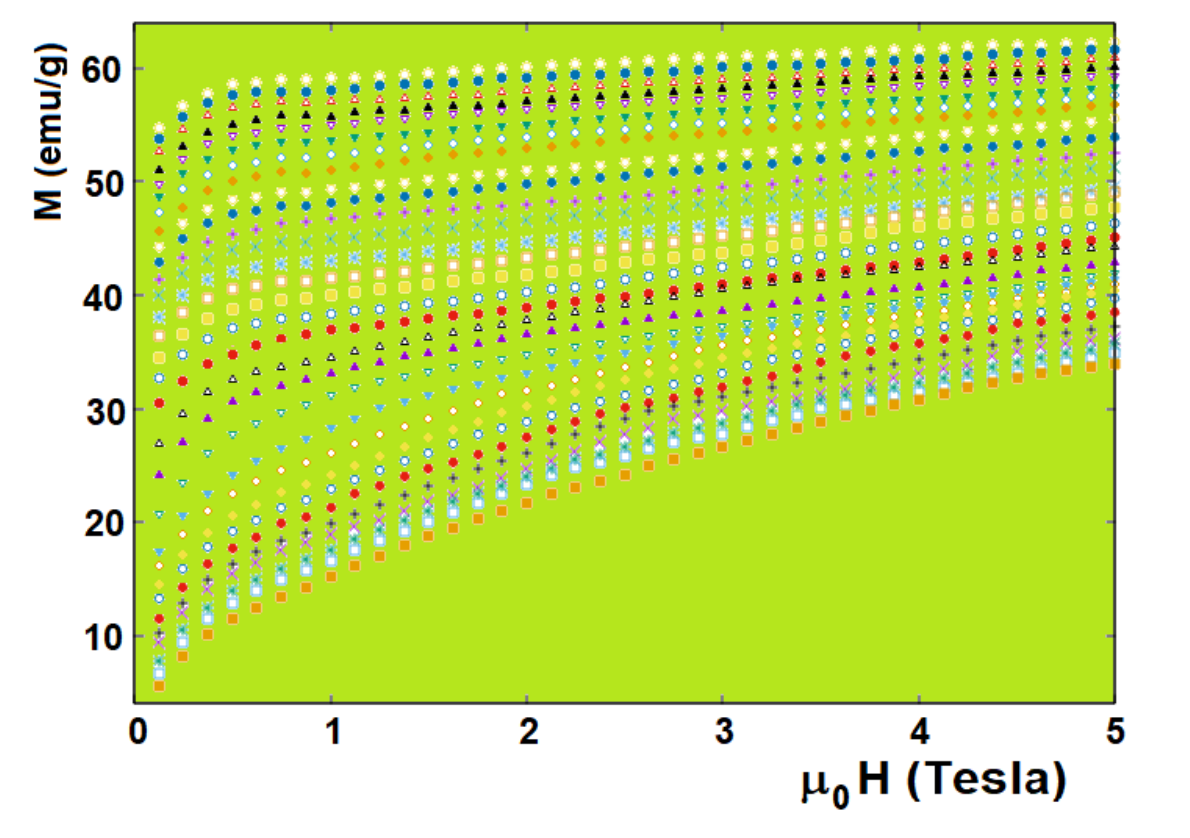}
\caption{Effect of magnetic field (Tesla) on the magnetization  numerically obtained for $T=260$ K, 262 K, ..., 320 K (from top to  bottom line).  }
\label{fig7}
\end{figure}

\begin{figure}[ht]
\centering
\includegraphics[width=10cm,angle=0]{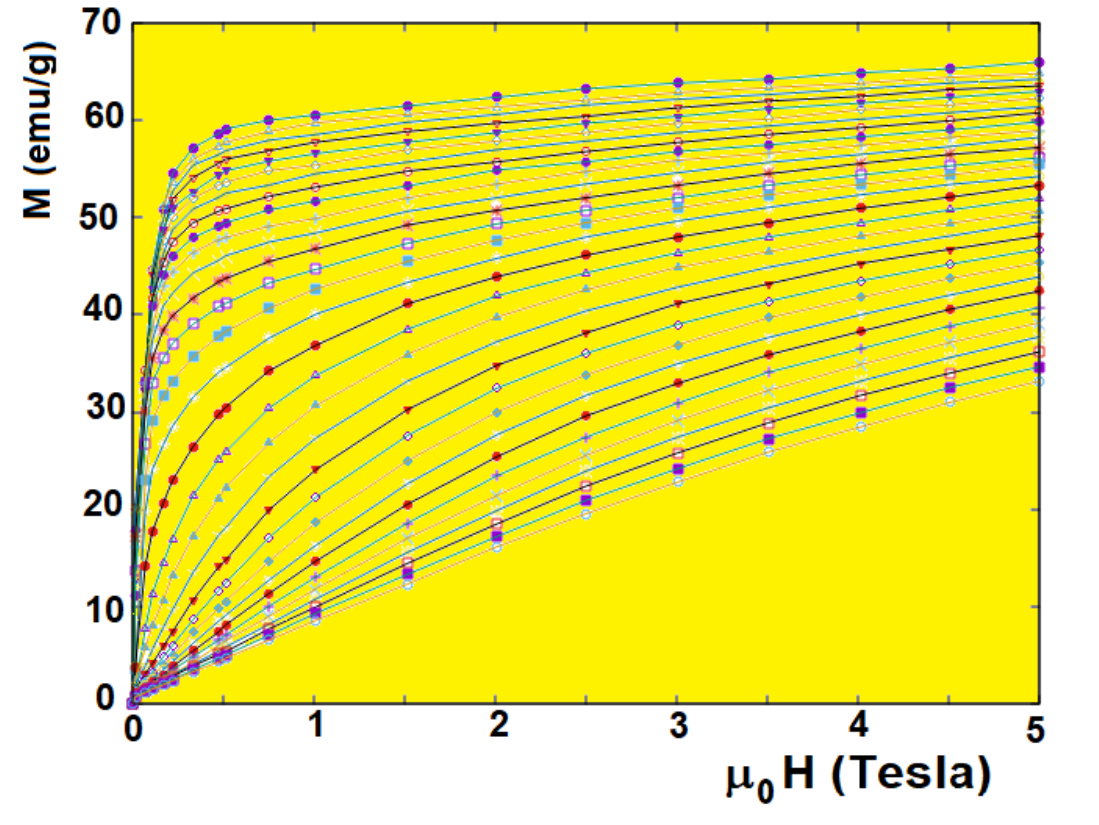}
\caption{Effect of magnetic field (Tesla) on the magnetization  experimentally obtained for $T=260$ K, 262 K, ..., 320 K (from top to  bottom line). This figure is made using data of Fig. 8 for sample P1200 of Ref. [23].  }
\label{fig8}
\end{figure}

Let us calculate the magnetic entropy change   $|\Delta S_m|$  defined by the Maxwell's formula 

\begin{equation}\label{DeltaS}
|\Delta S_m (T,H)|=\int_0^H\left[\frac{\delta M(T,H_i)}{\delta T} \right]_{H_i}\  \mu_0 dH_i
\end{equation}
where $\delta M$ is the variation of the magnetization at $H$ when $T$ varies from $T$ to $T+\delta T$.  This formula is discretized  and used in experiments and simulations as

\begin{equation}\label{DeltaS1}
|\Delta S_m (T,H)|=\sum_i\left[\frac{M_i-M_{i+1}}{T_{i+1}-T_i} \right]\mu_0 \Delta H_i
\end{equation}
The experimental magnetic entropy change $|\Delta S_m|$ using this formula   is shown in Fig. \ref{figDeltaS}a (sample P1200 of [23]): the peak temperature does not change significantly with increasing $H$ but the peak height is higher with larger $H$. For comparison $|\Delta S_m|$ obtained by MC simulations is  shown in Fig.\ref{figDeltaS}b.  We note that the peak heigth is slightly higher than the experimental one, for all $H$. We believe that this is a consequence of higher $M$ at very low $H$ mentioned above.

\begin{figure}[ht]
\centering.
\includegraphics[width=8cm,angle=0]{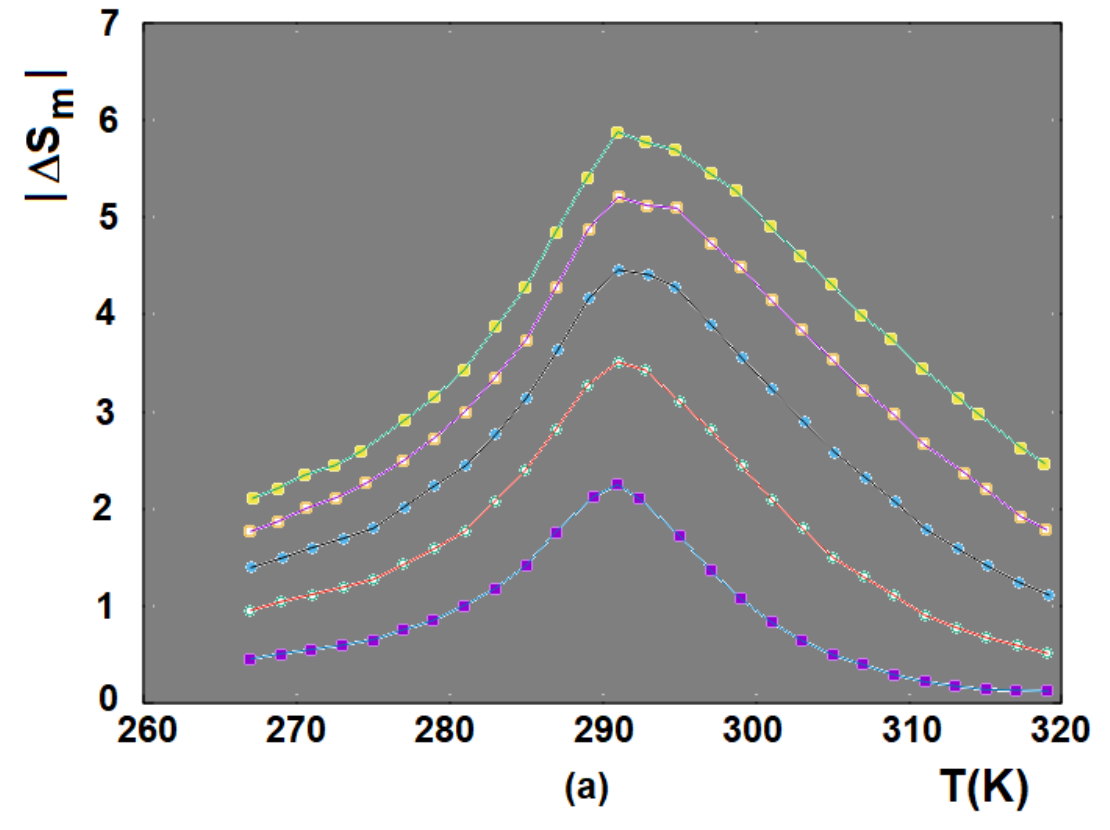}
\includegraphics[width=8cm,angle=0]{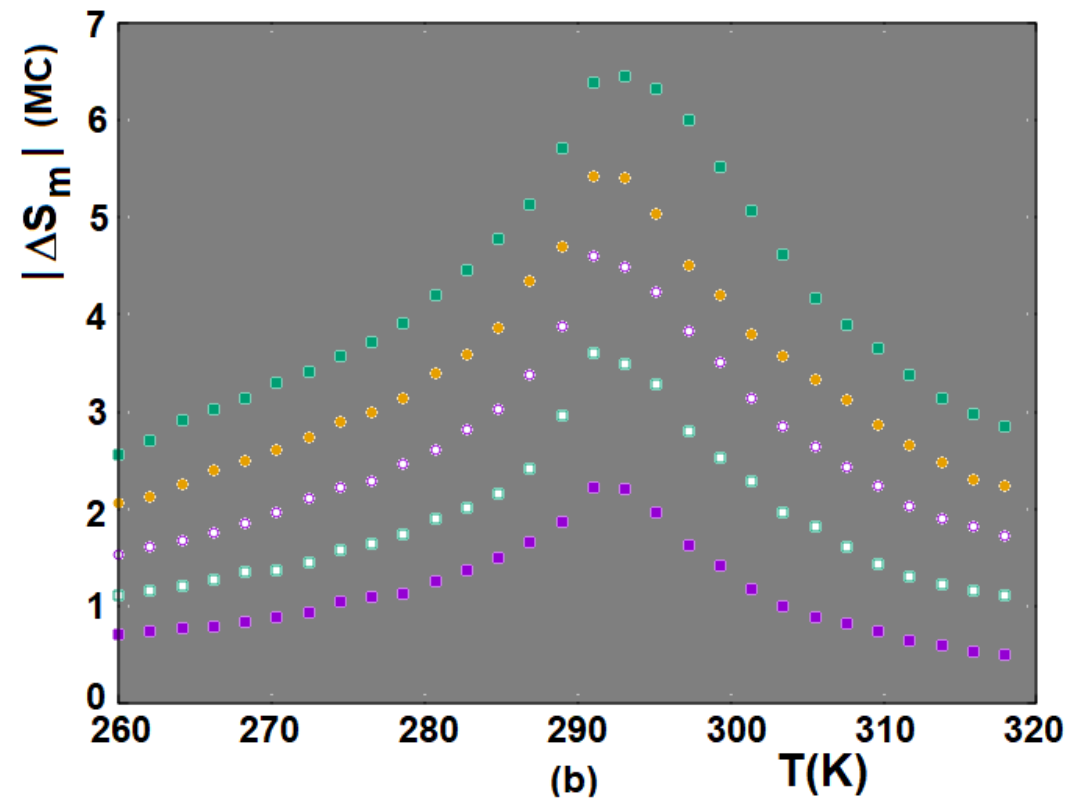}
\caption{(a)  Experimental magnetic entropy change $|\Delta S_m|$ in unit of J/(kgKelvin) versus  temperature $T$ for $\mu_0H=$1, 2, 3, 4, 5 Tesla (from bottom to top), results extracted from Ref. [23], (b)  $|\Delta S_m|$ [in unit of J/(kgKelvin)] obtained by MC simulations. See text for comment.}
\label{figDeltaS}
\end{figure}
Now, we calculate the Relative Cooling Power (RCP) using the following relation

\begin{equation}\label{RCP}
RCP(H)=|\Delta S_{max} (H)|\times \Delta T
\end{equation}
where $|\Delta S_{max} (H)|$ is the maximum value of $|\Delta S_{m} (H)|$ and  $\Delta T$  the temperature range at  the full
width at half maximum.  

We show in Table \ref{figRCP} the  experimental Relative Cooling Power (RCP) taken from Ref. [23] and our MC results of RCP.  Note that Ref. [23] gives only the RCP for 1 and 2 Tesla for the sample P1200. But using their curves shown in Fig. \ref{figDeltaS}, we calculated the experimental RCP for 3, 4 and 5 Tesla shown in the Table.  Our RCP values are sommewhat higher than the experimental values but they are  within the same order of magnitude. We consider this as a good agreement. 

%
\begin{table}
\centering
\caption{Experimental Relative Cooling Power and MC results of Pr$_{0.67}$Sr$_{0.33}$Mn$_{0.67}^{3+}$Mn$_{0.33}^{4+}$O$_{3}$, for  $\mu_0H$=1, 2, 3, 4, 5 Tesla. See text for comments.}
\label{figRCP}       
\begin{tabular}{lll}
\\
\hline
\\
$\mu_0H$(T) |& Experimental RCP(J/kg) |& MC RCP (J/kg)  \\
\\
\hline\\
1 & 36.4& 51.7 \\
2 & 79.2 & 88.8 \\
3& 121.5 & 130.8 \\
4&153.0&207.2\\
5&210.0&275.1\\
\\\hline
\end{tabular}
\end{table}

\section{Conclusion}\label{concl}

In this paper, we proposed a model Hamiltonian which includes a multi-spin interaction in addition to the pairwise interactions to study the magnetic properties of Pr$_{0.67}$Sr$_{0.33}$Mn$_{0.67}^{3+}$Mn$_{0.33}^{4+}$. We compared our results with experimental measurements of the magnetization $M$ versus temperature $T$ which shows that $M(T)$ unusually increases with increasing $T$ before making a vertical fall at the transition temperature $T_C$. 
With our model Hamiltonian, we performed numerical simulations and obtained results of $M(T)$ in good agreement with experiments for the whole range of temperature from 0 to $T_C$. In addition, we showed a clear evidence  of the first-order character of the transition, using the energy histogram technique.  

We also calculated the magnetic entropy change and the Relative Cooling Power (RCP) for applied fields from 1 to 5 Tesla. These results are in agrrement with experimental data. The high coefficients RCP make the present compound a potential candidate for magnetic refrigeration. 

We note that with the multi-spin interaction, we previously succeeded to obtain results for Pr$_{0.9}$Sr$_{0.1}$Mn$_{0.9}^{3+}$Mn$_{0.1}^{4+}$ [24] and Pr$_{0.55}$Sr$_{0.45}$Mn$_{0.55}^{3+}$Mn$_{0.45}^{4+}$ [25] which, in both cases, are in agreement with experiments. Rarely, modeling using a microscopic interaction Hamiltonian leads to good agreement on macroscopic properties experimentally observed.   

To conclude, we believe that the multi-spin interaction is necessary for studying  materials with complex structures such as in the above-mentioned compounds whose behaviors are unusual.   
\vspace{2cm}

\acknowledgments

Yethreb Essouda  is indebted to the CY Cergy Paris University for hospitality during her working visits.

\end{document}